\documentclass[useAMS,usenatbib]{mn2e}

\usepackage{lscape}
\usepackage{supertabular}
\usepackage{longtable}
\usepackage{graphicx}
\usepackage{txfonts}
\def\gtsima{$\; \buildrel > \over \sim \;$}
\def\ltsima{$\; \buildrel < \over \sim \;$}
\def\gsim{\lower.5ex\hbox{\gtsima}}
\def\lsim{\lower.5ex\hbox{\ltsima}}

\title[Disc variability in GX 339-4]     
  {The evolution of the disc variability along the hard state of the black hole transient GX~339-4}         
\author[B. De Marco et al]
  {B.~De Marco,$^{1}$\thanks{E-mail: bdemarco@mpe.mpg.de} 
  G.~Ponti,$^{1}$ 
  T.~Mu\~{n}oz-Darias,$^{2, 3, 4}$
  K.~Nandra,$^{1}$\\
  $^1$ Max-Planck-Institut f\"{u}r Extraterrestrische Physik, Giessenbachstrasse 1, D-85748, Garching, Germany\\
  $^2$ Instituto de Astrof\'isica de Canarias, 38205 La Laguna, Tenerife, Spain\\
  $^3$ Departamento de astrof\'isica, Univ. de La Laguna, E-38206 La Laguna, Tenerife, Spain\\
  $^4$ University of Oxford, Department of Physics, Astrophysics, Denys Wilkinson Building, Keble Road, Oxford, OX1 3RH, UK}
\date{Released 2014 Xxxxx XX}

\pagerange{\pageref{firstpage}--\pageref{lastpage}} \pubyear{2014}

\def\LaTeX{L\kern-.36em\raise.3ex\hbox{a}\kern-.15em
    T\kern-.1667em\lower.7ex\hbox{E}\kern-.125emX}

\begin{document}

\label{firstpage}

\maketitle

\begin{abstract}
We report on the analysis of hard-state power spectral density function (PSD) of GX 339-4 down to the \emph{soft} X-ray band, where the disc significantly contributes to the total emission.
At any luminosity probed, the disc in the hard state is intrinsically more variable than in the soft state.
However, the fast decrease of disc variability as a function of luminosity, combined with the increase of disc intensity,  causes a net drop of fractional variability at high luminosities and low energies, which reminds the well-known behaviour of disc-dominated energy bands in the soft state. 
The peak-frequency of the high-frequency Lorentzian (likely corresponding to the high-frequency break seen in active galactic nuclei, AGN) scales with luminosity, but we do not find evidence for a linear scaling. In addition, we observe that this characteristic frequency is energy-dependent.
We find that the normalization of the PSD at the peak of the high-frequency Lorentzian decreases with luminosity at all energies, though in the \emph{soft} band this trend is steeper. 
Together with the frequency shift, this yields quasi-constant \emph{high frequency} (5-20 Hz) fractional \emph{rms} at high energies, with less than 10 percent scatter. This reinforces previous claims suggesting that the \emph{high frequency} PSD solely  scales with BH mass.
On the other hand, this constancy breaks down in the \emph{soft} band (where the scatter increases to $\sim$ 30 percent). This is a consequence of the additional contribution from the disc component, and resembles the behaviour of optical variability in AGN.

\end{abstract}


\section{Introduction}
\label{intro}

Black hole X-ray binaries (BHXRB) spend most of their time in a quiescent state, which is occasionally interrupted by outbursts of activity. As the outburst evolves the source passes through different accretion states and both the variability and spectral properties change (e.g. Miyamoto et al. 1992; Belloni et al. 2005;  Remillard \& McClintock 2006). 
The variability properties of these sources have been extensively studied in the past, revealing the existence of trends as a function of accretion state and/or luminosity (e.g. Zdziarski et al. 2004; Mu\~noz-Darias, Motta, \& Belloni 2011; Belloni, Motta, \& Mu\~noz-Darias 2011). However, due to the sensitivity limits of the detectors employed (particularly the Proportional Counter Array, PCA, on the \emph{Rossi X-ray Timing Explorer}, RXTE), these studies have been mostly limited to energies $>$ 3 keV.
At these energies the \emph{broad-frequency-band} (e.g. in the typical frequency range $\sim0.1-60$ Hz) aperiodic variability progressively decreases as the outburst evolves, with the fractional root-mean-square (\emph{rms}) variability amplitude (e.g. Nandra et al. 1997; Vaughan et al 2003; Ponti et al. 2004) dropping from values of tens of percent in the canonical hard state, down to a few percent in the canonical soft state (Belloni et al. 2005; Mu\~{n}oz-Darias, Motta \& Belloni 2011).
During the hard state the fractional \emph{rms} shows an ``inverted spectrum'' (decreasing with energy), which switches to a ``hard spectrum'' (increasing with energy, at least above $\sim$5 keV) during intermediate and soft states (e.g. Zdziarski et al. 2004; Belloni et al. 2011, as illustrated in the inset of Fig. \ref{fig:rms}). 
These behaviours are usually ascribed to variations of the relative contribution of the main X-ray spectral components (e.g. Gierli\'nski \& Zdziarski 2005). Indeed, over the past years, the general idea was that the strong \emph{broad-frequency-band} variability characterizing the hard state was associated with the Comptonized hard X-ray emission (power law) from the corona (or the base of a radio-emitting jet, e.g. Zdziarski et al. 1998; Markoff, Nowak, \& Wilms 2005; Droulans et al. 2010), since this component dominates the X-ray spectrum during this state.
On the other hand, the disc-thermal component, known to be variable on the time scales of the outburst, was thought to be intrinsically constant on shorter (than about hours-minutes) time scales (e.g. Gierli\'nski \& Zdziarski 2005). Given that the disc dominates the spectrum up to energies of  $\sim 10$ keV during the soft state, it would then be responsible for the drop of \emph{broad-frequency-band} variability amplitude, and for the change of shape of the fractional \emph{rms} spectrum.\\
The way these variability properties extrapolate to soft X-ray energies ($<3$ keV) during the hard state is currently poorly known. Since, during the hard state the disc component gives non-negligible contribution in the soft X-ray energy band ($\lsim$ 1 keV; Tomsick et al. 2008; Kolehmainen, Done, \& D\'iaz-Trigo 2014), to some extent we would expect to observe behaviours similar to those characterizing the disc during the soft state.
However, the recent use of data from the EPIC-pn detector onboard XMM-Newton (sensitive down to $\sim$0.3 keV), for the spectral-timing analysis of BHXRBs, unveiled the existence of high levels of intrinsic (on time scales of $\gsim$1 s) disc variability in the hard state (Wilkinson \& Uttley, 2009). Though this result has been obtained studying a single dataset for each of the two sources GX 339-4 and Swift J1753.5-0127, it might represent a general property of BHXRBs. 
Thus, extending timing studies down to soft X-ray energies is crucial in order to gain a complete understanding of the disc-corona variability processes.\\
Another important property of X-ray variability in BHXRBs is that the characteristic time scales vary during the outburst. During the hard state all the variability components (commonly modeled with Lorentzians; Belloni, Psaltis, \& van der Klis 2002) in the power spectral density function (PSD) shift towards higher frequencies as the luminosity increases, while keeping tight correlations between each other (e.g. Belloni et al. 2005; Done \& Gierli\'nski 2005).
This indicates strong dependence of X-ray variability on the luminosity/mass-accretion rate of the source.
The shift of the characteristic frequencies as a function of mass-accretion rate is a general property of accreting systems. 
Indeed, as in BHXRBs, a scaling (consistent with being linear; McHardy et al. 2006; Koerding et al. 2007) also links the ``high-frequency break'' (Edelson \& Nandra 1999) and the mass-accretion rate in active galactic nuclei (AGN). However, despite all these variations in both characteristic frequencies and \emph{broad-frequency-band} variability amplitude, the \emph{high-frequency} PSD (i.e. above the peak-frequency of the high-frequency Lorentzian) remains remarkably constant during the entire outburst (Gierli\'{n}ski, Niko{\l}ajuk \& Czerny, 2008), indicating little or no-dependence on the luminosity and the accretion state of the source. 
In analogy with BHXRBs, also in AGN the high-frequency PSD (above the high-frequency break) does not show any clear dependence on the luminosity/mass-accretion rate (e.g. O'Neil et al. 2005; Zhou et al. 2010; Ponti et al. 2012).\\
In this paper we aim at investigating the evolution of the PSD as a function of luminosity/mass-accretion rate, extending the analysis down to the soft X-ray energy band. For this reason, we study the hard-state PSD of GX 339-4, the best monitored BHXRBs. We used data collected with XMM-Newton, and, in order to obtain a broad band coverage, we considered also simultaneous RXTE observations. The data selection and reduction are presented in Sect. \ref{reduction}, while the PSD estimate and analysis procedures are described in Sect. \ref{analysis}. Given the different, frequency-dependent behaviours outlined before, which characterize the PSD at energies $\gsim$ 3 keV, we will first study the \emph{broad-frequency-band} variability properties (Sect. \ref{sec:frms}-\ref{sec:pfreq}), and then more specifically focus on the \emph{high-frequency} part (Sect. \ref{sec:scaling}). Results are discussed in Sect. \ref{discussion}.

\begin{figure}
\centering
\vspace{2.5cm}
\begin{tabular}{p{10cm}}
\includegraphics[width=0.45\textwidth,angle=0]{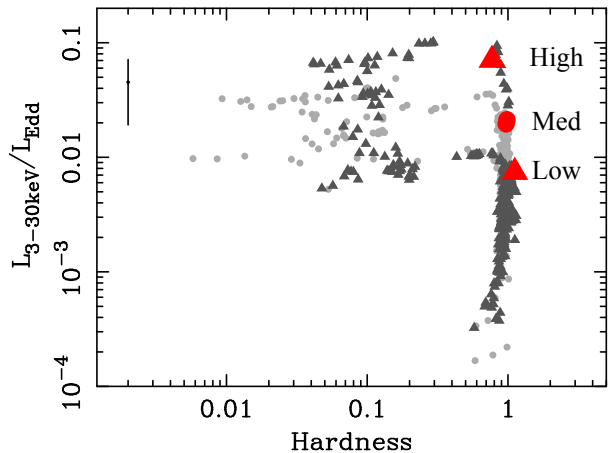}\\ 
\end{tabular}
\caption{The hardness-intensity diagram of GX 339-4 as obtained from all the archived RXTE observations relative to the 2004 (light grey dots) and 2009 (dark grey triangles) outbursts. The luminosity is computed in the range 3-30 keV, the hardness is computed as the ratio between the 13-30 keV and 3-13 keV flux. Colored points mark the position of the analyzed XMM-Newton and simultaneous RXTE observations in the diagram. The labels refer to the nomenclature used in the paper to indicate the 2009 (low luminosity), 2004 (medium luminosity), and 2010 (high luminosity) observations. The error bar on the left side of the plot reports the uncertainty on the HID normalization due to the uncertainties on the distance and the BH mass of GX 339-4.}
\label{fig:hd2}
\end{figure}

\section{Data reduction}
\label{reduction}
GX 339-4 has been observed 18 times by XMM-Newton (as of May 2014). In this paper we focus on all the archived observations with exposures $>$10 ks, which caught the source in a hard state, i.e four observations in total. In order to further increase the spectral band coverage we considered also all the available simultaneous RXTE observations. Details about the analysed data sets are reported in Table \ref{tab:ObsLog}.

\begin{table*}
\caption{Log of the analyzed XMM-Newton and simultaneous RXTE observations: (1) observation ID; (2) date; (3) detector used for the analysis and corresponding observing mode (for the RXTE observations the Proportional Counter Units  -- PCU -- coadded in the analysis are also reported);  (4) effective exposure (the XMM-Newton effective exposures refer to the duration of the observation after removal of high soft proton background, the percentage of this exposure affected by telemetry drop outs is reported within parentheses) (5) mean count rates in the energy bands used for the analysis (i.e. bands 0.5-1.5 keV -- \emph{soft} -- and 2-9 keV -- \emph{hard} -- for XMM-Newton, and band 10-30 keV -- \emph{very hard} -- for RXTE).}
\label{tab:ObsLog}
\centering
\vspace{0.2cm}

\begin{tabular}{c c c c c}
\hline
\multicolumn{5}{ |c| }{XMM-Newton} \\
\hline          

	(1)             & (2)  & (3) & (4) & (5) \\
    Obs ID          & Date & Instrument/Mode                   & Exp (drop outs)             &  Mean count rate (soft/hard)  \\
    
    			&  & &  [s] & [cts/s] \\

\hline

   0204730201  &  2004-03-16  &   Epic-PN/Timing    & \multicolumn{1}{ l }{104200 (20.7 percent)}  & 90/108 \\ 
   0204730301  & 2004-03-18  &  Epic-PN/Timing     &  \multicolumn{1}{ l }{71000  (21 percent)} & 98/117 \\
   0605610201 & 2009-03-26  &  Epic-PN/Timing     &  \multicolumn{1}{ l }{32220 (0.2 percent)} & 43/61 \\
   0654130401   &  2010-03-28  &  Epic-PN/Timing  &  \multicolumn{1}{ l }{33630 (23.7 percent)} & 357/434  \\

\hline
\multicolumn{5}{ |c| }{RXTE} \\
\hline          

    Obs ID          & Date & Instrument/Mode                   & Exp  &  Mean count rate (very hard)  \\
    
        			&  & &  [s] & [cts/s/PCU]  \\
\hline
  90118-01-05-00    &  2004-03-16  &  \multicolumn{1}{ l }{PCU 0,2,3/Good Xenon} & 2324 &  55   \\ 
  90118-01-06-00   & 2004-03-17 &  \multicolumn{1}{ l }{PCU 0,2/Good Xenon}    & 1660  & 58  \\ 
  90118-01-07-00   & 2004-03-18  &    \multicolumn{1}{ l }{PCU 0,2/Good Xenon}   &  6505  & 60    \\ 
  94405-01-03-00  & 2009-03-26  & \multicolumn{1}{ l }{PCU 0,2/Good Xenon} & 3055 & 38  \\
  94405-01-03-01  &  2009-03-26 & \multicolumn{1}{ l }{PCU 2/Good Xenon} & 3305 & 37 \\
  94405-01-03-02  & 2009-03-26 & \multicolumn{1}{ l }{PCU 1,2/Good Xenon} & 1807 & 39   \\
  95409-01-12-01   &  2010-03-28  &  \multicolumn{1}{ l }{PCU 2/Generic Event} &  19195  & 180   \\ 
\hline

\end{tabular}

\end{table*}

Among the four selected XMM-Newton observations, two (16-18 March 2004) were carried out during consecutive satellite revolutions. Hereafter, we will show results obtained by combining these two observations together, and we will refer to them as one single observation.
The remaining two observations belong to the same outburst of the source, but they have been carried out one year apart (in 2009 and 2010).\\
We used EPIC-pn (Str\"uder et al. 2001) and Proportional Counter Array (PCA; Jahoda et al. 2006) data, which ensure high effective area over a broad spectral range.\\
The XMM-Newton data reduction was done using the XMM Science Analysis System (SAS v13.5), and the latest calibration files (CCF, as of May 2014). The analyzed data sets are all in Timing mode. We followed standard reduction procedures, and applied Rate Dependent PHA (RDPHA) and X-ray loading (XRL) corrections (Guainazzi \& Smith 2013; Guainazzi 2014, and references therein). We selected time intervals free from background proton flares (GTI, the final effective exposures are listed in Table \ref{tab:ObsLog}) and considered only events with PATTERN $\leq4$. The source counts were extracted from the region which includes 90 percent of the total collected counts. For all the observations this region falls between RAWX columns 31 and 45. \\
Being all the observations in Timing mode, pile-up is expected to be an issue only when the count rate exceeds $\sim$800 cts/s (in the 0.7-10 keV range, Guainazzi et al. 2014). This threshold is well above the average count rate registered during the 2004 and 2009 observations considered here. However, the 2010 observation exceeds this threshold. In addition, variability may cause pile-up during short time intervals even when the average count rate is below the threshold.
Therefore we verified the pile-up level during each observation using the SAS tool ``epatplot''. In the 2010 and 2004 observations, the observed \emph{single} and \emph{double} events fractions slightly deviate from the expected trend for data not affected by pile-up (respectively by $\sim$1.2 and 12 percent in the 2-10 keV band for both observations), but this deviation is systematic at all energies. As noticed in Done \& Diaz-Trigo (2010), these systematic deviations might be either due to the uncertainties related to the modeling of the \emph{singles} and \emph{doubles} fraction for the Timing mode, or might indicate that the data are actually slightly affected by pile-up. 
Thus, we tested our results against pile-up by repeating the analysis (described in Sect. \ref{analysis}) of the 2010 observation (which has the highest EPIC-pn count rate) on the event file obtained by excising the central three pixels (i.e. RAWX$=37-39$), which reduces the excess of \emph{double} events due to pile-up by a factor $\sim$5 in the 2-10 keV band. We found results perfectly consistent with those presented in Sect. \ref{results}, the only difference being the expected reduction of variability power signal-to-noise (S/N) ratio in the PSD.\\
For the RXTE data analysis we followed standard procedures\footnote{The reduction procedures are described in the RXTE data reduction cookbook: http://heasarc.nasa.gov/docs/xte/recipes/cook\_book.html} using the HEASOFT software v6.13. For the timing analysis, we used either Good Xenon or Generic Event configuration modes, depending on the availability (see Table \ref{tab:ObsLog}). 
To maximize the S/N ratio we extracted data from all the layers of the Proportional Counter Units (PCU) that were simultaneously and continuously switched on during each observation. On the other hand, the luminosities are computed from data in Standard2 configuration, considering only the top layer of PCU2.\\ 
The position of the analyzed observations within the hardness-intensity diagram (HID) is displayed in Fig. \ref{fig:hd2}. This plot includes all the RXTE data relative to the 2004 (light grey dots) and 2009 (dark grey triangles) outbursts. Fluxes are computed in the 3-30 keV energy range, assuming a simple power law plus a gaussian at the energy of the Fe K$\alpha$ line. The hardness ratio is computed as the ratio between the 3-13 keV and 13-30 keV fluxes. The fluxes are converted into Eddington-scaled luminosities. Eddington luminosities are obtained by considering the currently available estimates and uncertainties on the distance and the BH mass of the source (i.e. $\sim$6-10 kpc, and $\sim$6-10 M$_{\odot}$, respectively, Hynes et al. 2003, 2004, Mu\~{n}oz-Darias et al. 2008) and assuming a reasonable mean value, i.e. 8 kpc and 8 M$_{\odot}$. 
The error bar displayed on the left side of the plot highlights the uncertainty on the plot normalization, as due to the uncertainties on the distance and the BH mass. Note that, when considering the relative increase of luminosity, the relevant uncertainty is the error on the flux measurements, which however is very small (i.e. of the order of $\sim 0.1-0.3$ percent), thus it has not been reported in the plot.
 The analyzed observations show luminosities $L_{3-30keV}/L_{Edd} \sim$$0.007$ (2009 observations), $\sim$$0.02$ (2004 observations), and $\sim$$0.07$ (2010 observations). In the following we will refer to these observations respectively as ``low'', ``medium'', and ``high luminosity''.

\begin{table}
\caption{Results obtained from the fit of the PSDs of GX 339-4 with a model comprising three broad Lorentzians. Best-fit values of the peak-frequencies (in units of Hz) of the lowest, intermediate, and highest frequency Lorentzian (respectively, $\nu_{l}$, $\nu_{i}$, and $\nu_{h}$) are reported, in the \emph{soft} (0.5-1.5 keV, S), \emph{hard} (2-9 keV, H), and \emph{very hard} (10-30 keV, VH) band, and at different luminosities. Errors are reported at 90 percent confidence level.}
\label{tab:fits}
\centering
\vspace{0.2cm}
\begin{tabular}{c c c c}
\hline       
  Band     &  Low L  &  Med L  & High L \\  
   
               &  \multicolumn{3}{ |c| }{$\nu_{l}$}  \\ 
               
\hline
S   & 0.0096 $\pm$ 0.0014 &   0.042$\pm$0.002   &  0.146$\pm$0.013  \\
H   &  0.0092 $\pm$ 0.0014 &   0.043$\pm$ 0.003  &  0.172$\pm$0.026  \\
VH &  0.0080 $\pm$ 0.0035 &    0.043$\pm$0.004   & 0.179$\pm$0.016  \\
\hline
               &  \multicolumn{3}{ |c| }{$\nu_{i}$}     \\ 
\hline
S   &  0.18$\pm$0.04 &  0.56$\pm$0.05   & 0.166$\pm$0.008 \\
H   &   0.22$\pm$0.02 &  0.76$\pm$0.05   &  0.161$\pm$0.008 \\
VH &  0.17$\pm$0.06 &   0.55$\pm$0.08  &  0.161$\pm$0.012 \\
\hline
               &  \multicolumn{3}{ |c| }{$\nu_{h}$}     \\ 
\hline
S   &  1.15$\pm$0.15       &  1.18$\pm$0.10   & 1.85$\pm$0.15  \\
H   &  1.88$\pm$0.15     &  2.79$\pm$0.29     & 2.89$\pm$0.11  \\
VH &  1.90$\pm$0.42        & 2.58$\pm$0.44      & 3.05$\pm$0.28 \\
\hline
                & \multicolumn{3}{ |c| }{$\chi^2$/dof}  \\  
\hline
S              & 80.6/42 & 161.9/66  & 21.4/25  \\  
H              & 95.9/42  & 87.1/66  & 60.3/25  \\  
VH              & 44.4/42  & 89.4/30 & 36.8/25 \\  

\hline
\end{tabular}

\end{table}

\begin{figure}
\centering
\vspace{1.cm}
\begin{tabular}{p{7.6cm}}
\includegraphics[height=8cm,angle=270]{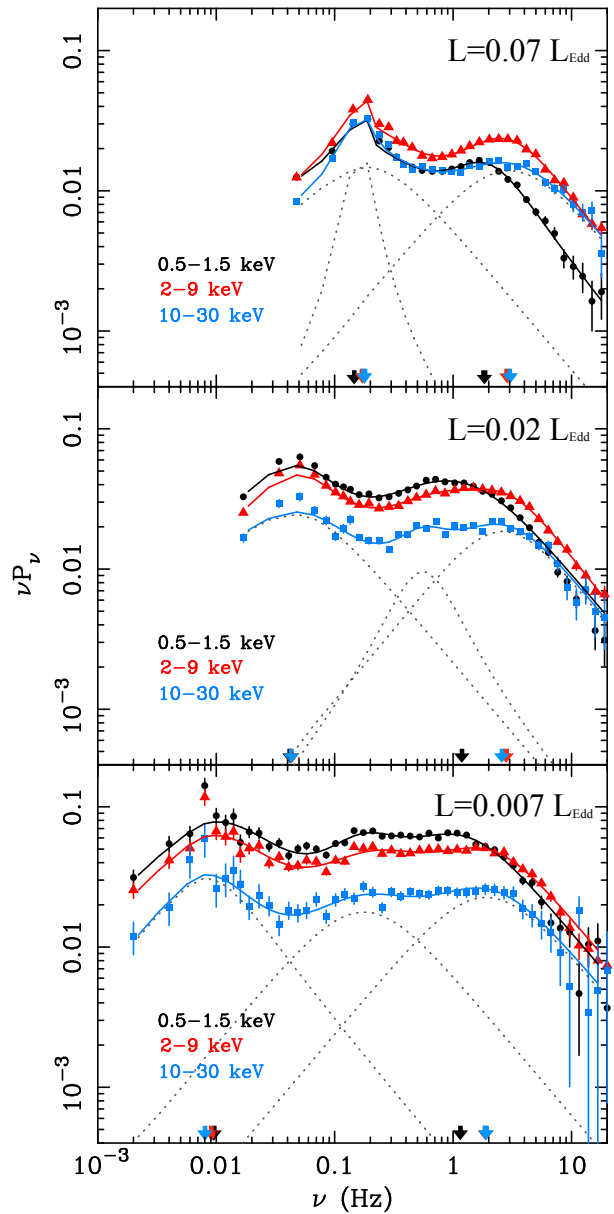}
\end{tabular}
\caption{The PSDs of GX 339-4 in the three energy bands 0.5-1.5 keV (\emph{soft}), 2-9 keV (\emph{hard}), and 10-30 keV (\emph{very hard}), for each observation at different luminosity. The arrows mark the best-fit peak frequency of the low-frequency ($\nu_l$) and the high-frequency ($\nu_h$) Lorentzians for each energy band.}
\label{fig:PSD1}
\end{figure}

\begin{figure}
\centering
\vspace{1.cm}
\begin{tabular}{p{7.6cm}}
\includegraphics[height=8cm,angle=270]{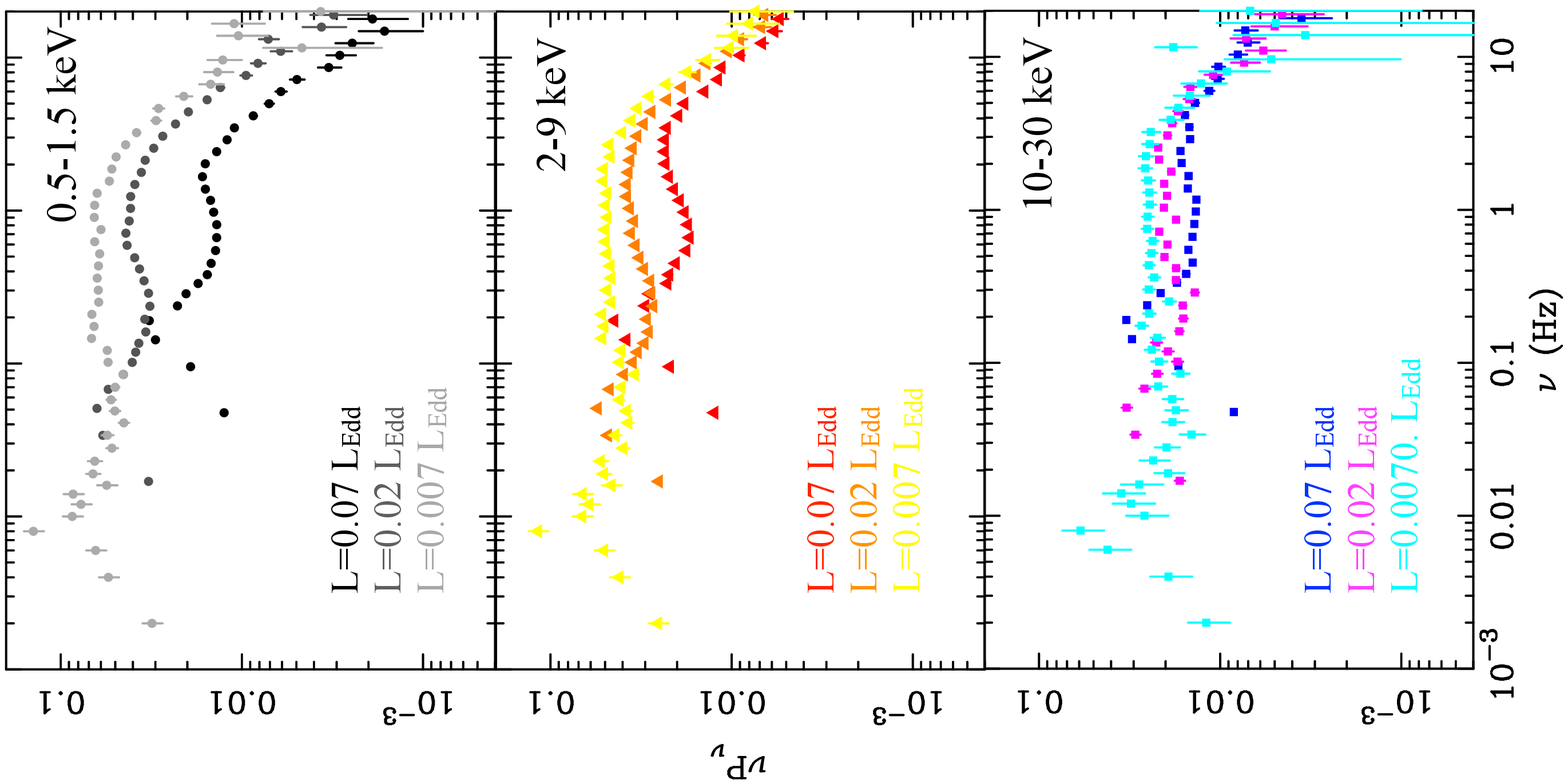}
\end{tabular}
\caption{Overplotted are the PSDs of GX 339-4 at different luminosities, for each of the energy bands \emph{soft} (upper panel), \emph{hard} (middle panel), and \emph{very hard} (lower panel).}
\label{fig:PSD2}
\end{figure}

\section{Data analysis}
\label{analysis}
For each observation we extracted light curves with time bins of 1ms (we verified that this choice of the time bin does not cause any spurious feature within the analysed frequency window, which might affect the analysis and results presented in this paper) in the three energy bands: 0.5-1.5 keV, 2-9 keV (using XMM-Newton data), and 10-30 keV (using RXTE data). Hereafter we will refer to these bands respectively as \emph{soft}, \emph{hard}, and \emph{very hard}. 
According to previous analyses of the same data sets (Kolehmainen et al. 2014; Plant et al. 2015), the chosen \emph{soft} band is where the accretion disc gives the most significant contribution. In the other two energy bands the dominant component is the primary hard X-ray power law (this is shown also in Sect. \ref{sec:spec}).\\
To study the evolution of the variability characteristics of GX 339-4 at different energies, we estimated the PSD within each of these energy bands. To this aim, the light curves were sampled over intervals shorter than the total exposure and of equal length within each observation. The length of the segments for the XMM-Newton light curves was chosen so as to exclude the gaps due to telemetry drop-outs (which occur mostly when the target source is very bright, i.e. during the high and medium luminosity observations), while ensuring a sufficiently broad frequency coverage. For consistency, the RXTE light curves have been sampled according to the corresponding simultaneous XMM-Newton light curves (i.e. 21 s, 59 s, and 500 s for the high, medium and low luminosity observations, respectively). 
We estimated the PSD (e.g. Nowak et al. 1999b) of each light curve segment, using routines implemented with IDL v6.4, and averaged them to obtain an estimate of the intrinsic PSD for every observation and energy band. The level of variability power due to Poisson noise was determined fitting the high-frequency part (above 100 Hz) of the PSD with a constant. 
Then, the average PSDs have been Poisson noise-subtracted, and renormalized adopting the squared fractional rms normalization (Miyamoto et al 1991). We corrected the RXTE data for instrumental dead-time (which produces an over-estimate of the intrinsic Poisson noise level) using the dead-time model of Zhang et al. (1995, 1996). Since the variability power S/N ratio drops at high frequencies, we limited the analysis to frequencies $\lsim50$ Hz.\\
The hard-state PSDs of BHXRBs are usually very complex. They contain several structures which can be modeled with multiple Lorentzian components (e.g. Nowak 2000, Belloni et al. 2002, Pottschmidt et al 2003). 
We followed this same approach for the modeling of the PSDs of GX 339-4. 

\section{Results}
\label{results}

The hard-state PSDs of GX 339-4 are shown in Figs. \ref{fig:PSD1} and \ref{fig:PSD2}. In Fig. \ref{fig:PSD1} we show for each observation, the PSDs at different energies, while in Fig. \ref{fig:PSD2} we show for each energy band, the PSDs at different luminosities.\\
All the PSDs show a similar shape, with two broad humps, whose intensity and characteristic frequencies clearly change as a function of both luminosity and energy. These two broad humps are a characteristic feature of BHXRBs in the hard state (e.g. Nowak, Wilms, \& Dove 1999a; Wilkinson \& Uttley 2009; B\"{o}ck et al 2011; Grinberg et al. 2014; Cassatella et al. 2012). 
At frequencies lower than those covered here the variability power per unit logarithmic interval ($\nu P_{\nu}$) is expected to drop in the hard state (\emph{band-limited} noise). The chosen length of light curve segments allows us to obtain a good sampling of the broad-band noise of GX 339-4, down to the onset of this drop. We verified the absence of additional low-frequency broad structures by inspecting the PSDs obtained using longer segments of RXTE data. This confirmed that the analysed PSDs are band-limited, at least in the \emph{hard} and \emph{very hard} bands. We note that the \emph{soft} band PSD shows the same structures as those seen at higher energies, therefore we assume that also the \emph{soft} band PSD is band-limited.\\
We used two Lorentzian components to model the two humps and more rigorously quantify their variations. A third Lorentzian is required in all the fits to account for residual structures\footnote{The third Lorentzian fits the narrow peak of the low-frequency broad-band noise component of the high luminosity observation. In the other two observations, this Lorentzian fits an additional broad component at intermediate frequencies (see Table \ref{tab:fits}).}. 
The best-fit centroid frequency and full width at half maximum have been used to estimate the peak-frequency (i.e. the frequency of the maximum power per logarithmic interval; Belloni et al. 2002) of each Lorentzian (see Table \ref{tab:fits}) using the formula reported in Sect. \ref{sec:pfreq}.
The quality of the data is high enough, that our simple model, though providing a good description of each PSD, does not always return sufficiently small values of the reduced $\chi^{2}$ (see Table \ref{tab:fits}).
We verified that this is mainly due to the presence of small residuals which are left unmodeled. These residuals are usually accounted for by increasing the number of Lorentzian components (e.g. Nowak 2000, Belloni et al 2005). However, comparing our results with those found in the literature, both for GX 339-4 (e.g. Nowak at al. 1999a, Nowak 2000) and other BHXRBs (e.g. Cassatella et al. 2012, Grinberg et al. 2014) we conclude that three variability components are sufficient to obtain a good description of the hard-state broad-band noise of GX 339-4 within the analysed frequency window (i.e. up to frequencies of 20 Hz). Our best-fit models are plotted in Fig. \ref{fig:PSD1} (solid lines) together with the single components (dotted curves). To avoid confusion, only the components associated with the \emph{very hard} band PSD are shown in the plots. These best-fit PSD models were first used to study the variations of the \emph{broad-frequency-band} (between 0.002-20 Hz) fractional \emph{rms} variability amplitude (Sect. \ref{sec:frms}). Then we used them to constrain the characteristic frequencies and study their variations as a function of energy and luminosity (Sect. \ref{sec:pfreq}). Finally, we focused on the variability properties of the \emph{high-frequency} (between 5-20 Hz) portion of the PSD (Sect. \ref{sec:scaling}).

\begin{table}
\caption{The fractional \emph{rms} (e.g. Vaughan et al 2003) as obtained integrating the PSD best-fit models of each observation at different luminosities, and in each of the three bands 0.5-1.5 keV (\emph{soft}), 2-9 keV (\emph{hard}), 10-30 keV (\emph{very hard}). The \emph{broad-frequency-band} fractional \emph{rms} and the fractional \emph{rms} associated with each Lorentzian are obtained integrating over the frequency interval 0.002-20 Hz. Finally we report the \emph{high-frequency} fractional \emph{rms} obtained integrating over the interval $5-20$ Hz. Errors are computed at the 90 percent confidence level.}
\label{tab:rms}
\centering
\vspace{0.2cm}

\begin{tabular}{c c c c}
\hline          

              & \emph{Soft}                  & \emph{Hard}                 &  \emph{Very Hard} \\
\hline
\multicolumn{4}{ |c| }{Broad-frequency-band} \\
\hline
High L   & 0.318$\pm$0.022    &  0.376$\pm$0.024 &  0.318$\pm$0.020   \\ 
Med L   & 0.492$\pm$0.034    & 0.496$\pm$0.031 &  0.374$\pm$0.024   \\
Low L   & 0.670$\pm$0.036    & 0.600$\pm$0.036 & 0.433$\pm$0.024   \\ 
\hline
\multicolumn{4}{ |c| }{Low-frequency Lorentzian} \\
\hline
High L   & 0.231$\pm$0.024  &   0.251$\pm$0.026 & 0.203$\pm$0.022  \\ 
Med L   &  0.363 $\pm$0.037  & 0.349$\pm$0.035     & 0.261$\pm$0.027 \\
Low L   &  0.426 $\pm$0.046  & 0.384 $\pm$0.041 & 0.281$\pm$0.023 \\
\hline
\multicolumn{4}{ |c| }{High-frequency Lorentzian} \\
\hline
High L   & 0.187$\pm$0.020  & 0.223$\pm$0.022  &   0.189$\pm$0.019 \\
Med L   & 0.312$\pm$0.032 & 0.240$\pm$0.028  & 0.223$\pm$0.021 \\
Low L   & 0.386 $\pm$0.030 & 0.335$\pm$0.031  &  0.245$\pm$0.026 \\
\hline
\multicolumn{4}{ |c| }{5-20 Hz} \\
\hline
High L   &  0.075$\pm$0.007 &  0.118$\pm$0.011 & 0.103$\pm$0.010 \\ 
Med L   & 0.109$\pm$0.010 &  0.138$\pm$ 0.011 &  0.112$\pm$0.011 \\
Low L   &  0.132$\pm$ 0.010 &  0.147$\pm$ 0.012 & 0.108$\pm$0.010 \\
\hline
\end{tabular}

\end{table}

\subsection{Trends of \emph{broad-frequency-band} fractional \emph{rms} as a function of energy and luminosity}
\label{sec:frms}

\begin{figure}
\centering
\vspace{1.5cm}
\begin{tabular}{p{8cm}}

\includegraphics[width=0.45\textwidth,angle=0]{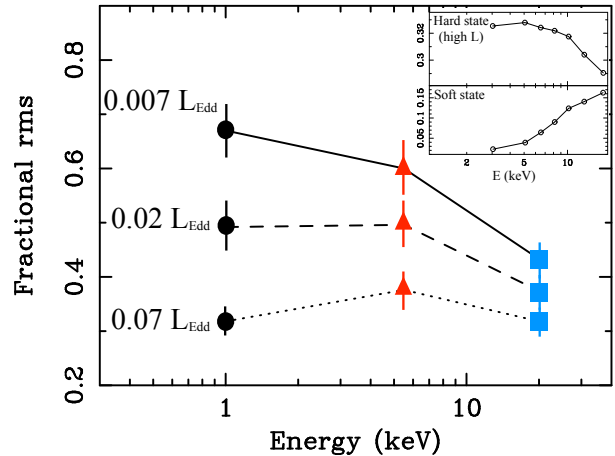} 

 \end{tabular}
\caption{Variations of the \emph{broad-frequency-band} fractional \emph{rms} spectrum with luminosity. Different symbols refer to the energy bands: \emph{soft} (black dots), \emph{hard} (red triangles), and \emph{very hard} (blue squares). The inset shows the typical spectral shape of the 0.1-32 Hz fractional \emph{rms} of GX 339-4 during a (high luminosity) hard and soft state observation in the RXTE bandpass (adapted from Belloni et al. 2011).}
\label{fig:rms}

\end{figure}

We integrated the best-fit PSD models for each luminosity and energy band, over the frequency interval 0.002-20 Hz (\emph{broad-frequency-band} fractional \emph{rms}), respectively for the high, medium, and low luminosity observations. Note that there is no evidence of additional low-frequency broad noise structures in the RXTE data (Sect. \ref{results}), and therefore these best-fit models describe the data well down to 0.002 Hz, at least in the \emph{hard} and \emph{very hard} bands. Given that the \emph{soft} band PSD shows similar structures as observed at higher energies, we assume that the extrapolation of our best-fit model describes the data well down to very low frequencies also in this band. We also estimated the fractional \emph{rms} associated with the low- and high-frequency Lorentzian (obtained integrating the single components of the best-fit PSD models within the same \emph{broad-frequency-band} interval). 
Results are reported in Table \ref{tab:rms}.
The \emph{broad-frequency-band} fractional \emph{rms}, independently of luminosity and energy band, is always high ($\gsim$30 percent), covering values typical of the hard state (Mu\~{n}oz-Darias et al 2011). 
In Fig. \ref{fig:rms} we plot the \emph{broad-frequency-band} fractional \emph{rms} as a function of energy (within the three energy bands adopted throughout this paper). 
Note that the fractional \emph{rms} of each Lorentzian shows the same energy-dependence as the \emph{broad-frequency-band} fractional \emph{rms}. For comparison we also report (inset of Fig. \ref{fig:rms}) the spectral shape of the 0.1-32 Hz fractional \emph{rms} of GX 339-4 during a (high luminosity) hard and the soft state, at the energies covered by the RXTE PCA (adapted from Belloni et al. 2011).\\
Our main results can be summarized as follows:
\begin{itemize}
\item  Above $\sim$2 keV, where the power law component dominates the source emission, we always observe a decrease of \emph{broad-frequency-band} fractional \emph{rms} as a function of energy. Therefore such a decrease of variability with energy appears to be an intrinsic property of the power law component. This might be produced by variations of the spectral index, with a high-energy pivot point. Extrapolating this trend onto the \emph{soft} energy band, we see that the \emph{broad-frequency-band} fractional \emph{rms} in this energy band is lower than expected. This is consistent with the presence of a disc-component, less variable than the power law.

\item Comparing the decrease of \emph{broad-frequency-band} fractional \emph{rms} within each energy band as a function of luminosity, we find that the largest drop is observed in the \emph{soft} energy band (a factor $\sim$2.1, as compared to $\sim$1.6 and 1.4 in the \emph{hard} and \emph{very hard} energy bands). In Sect. \ref{sec:spec}, we show that, though the intrinsic fractional \emph{rms} of the disc-component is high in the hard state, the fraction of variable disc flux in the \emph{soft} band decreases fast (faster than the power law component), thus causing the observed drop of \emph{soft} band variability as the luminosity increases.

\item In both the medium and low luminosity observations, the \emph{broad-frequency-band} fractional \emph{rms} spectrum has an overall ``inverted'' shape (decreasing with energy), typical of the hard state (e.g. Zdziarski et al. 2004; Belloni et al. 2011; inset of Fig. \ref{fig:rms}). However, during the high luminosity observation, the large drop of \emph{broad-frequency-band} fractional \emph{rms} observed in the \emph{soft} band causes a shift of the peak of variability power towards intermediate energies (i.e. in the \emph{hard} band). This appears to be the onset of a transition towards a ``hard spectrum'', typical of intermediate and soft states (inset of Fig. \ref{fig:rms}).

\end{itemize}

In order to characterize the nature of the drop of \emph{broad-frequency-band} fractional \emph{rms} observed in the \emph{soft} energy band at high luminosity, and investigate how this can be associated with an increasing contribution of disc-thermal emission in this band, we carried out fits of the XMM-Newton energy and covariance spectra (e.g. Wilkinson \& Uttley 2009; Uttley et al. 2011; Uttley et al. 2014).

\begin{figure}
\centering
\vspace{1.5cm}
\begin{tabular}{p{8.8cm}}

\includegraphics[width=0.45\textwidth,angle=0]{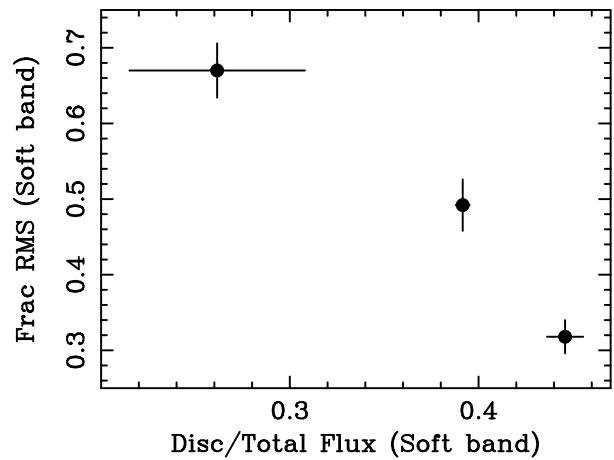} 

 \end{tabular}
\caption{The \emph{broad-frequency-band} fractional \emph{rms} as a function of disc-to-total (absorbed) flux ratio in the \emph{soft} band.}
\label{fig:rms_vs_disc}

\end{figure}

\subsubsection{Is the disc variable or constant during the hard state?}
\label{sec:spec}
\begin{figure*}
\centering
\vspace{1.5cm}
\begin{tabular}{p{8.8cm}}

\includegraphics[width=0.75\textwidth,angle=0]{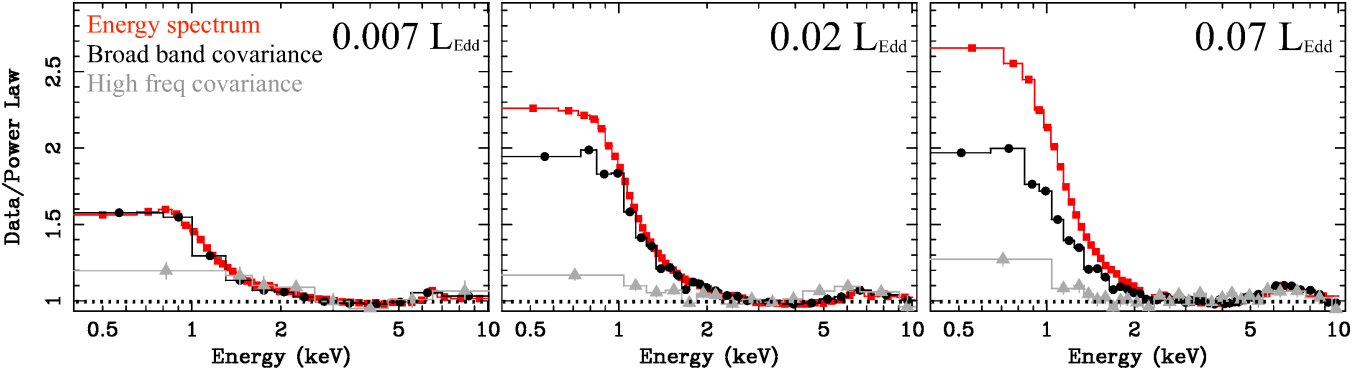} 

 \end{tabular}
\caption{Data-to-model ratios of the energy (red squares) and covariance spectra of the low (\emph{left panel}), medium (\emph{middle panel}), and high luminosity (\emph{right panel}) observations. The model used to fit the data is $tbabs*[diskbb+nthcomp]$, and the normalization of the $diskbb$ component has been set to zero to make the plots. All the covariance spectra have been computed using the 0.5-10 keV range as the reference band. The \emph{broad-frequency-band} covariance spectra (black circles) are obtained integrating over the entire frequency range covered by each data set up to frequencies of 20 Hz (i.e. 0.002-20 Hz, 0.017-20 Hz, and 0.04-20 Hz respectively for the low, medium, and high luminosity observations). The \emph{high-frequency} covariance spectra (light grey triangles) are obtained integrating over the frequency range 5-20 Hz.}
\label{fig:covar}
\end{figure*}

Previous analyses of the same (or part of the) data sets analysed in this paper (Wilkinson \& Uttley 2009; Kolehmainen et al. 2014) revealed the presence of a significant contribution from the accretion disc-thermal emission in the \emph{soft} band. Thus we used the XMM-Newton data to measure the disc-thermal emission-to-total flux ratio in the \emph{soft} band and verify whether the drop of fractional \emph{rms} is related to spectral variations in this band.
We fit the XMM-Newton spectra between 0.5-10 keV using a simple model for the continuum (i.e. $tbabs*[diskbb+nthcomp]$, with the Comptonization high energy cut-off fixed at 100 keV and the seed photon temperature tied to the disc temperature), and excluding the range of energies $4-7$ keV, dominated by the Fe K$\alpha$ line. We also excluded the energies around the edges of the response matrix, since residuals at these energies might be an artifact of uncorrected XRL effects and/or charge transfer inefficiency (Kolehmainen et al. 2014). The contribution from the disc emission strongly depends on the cold absorption column density parameter, $N_{H}$. To be more conservative, we let this parameter free. Hence, all the uncertainties stated below account for the uncertainty on $N_{H}$. Nevertheless, the derived best-fit values range between $N_{H} \sim 5-6 \times 10^{21}$ cm$^{-2}$, in agreement with the expected values (Dickey \& Lockman 1990; Kong et al. 2000).
In Fig. \ref{fig:rms_vs_disc} we plot the \emph{broad-frequency-band} fractional \emph{rms} as a function of disc thermal emission-to-total (absorbed) flux ratio in the \emph{soft} band, for the different luminosities. A net increase of the disc relative contribution to the total flux as the luminosity increases is observed. This increase coincides with the net decrease of fractional \emph{rms}, and the relation between the two is consistent with being linear (the best fit slope being $-1.26\pm0.24$). Thus the observed drop of \emph{broad-frequency-band} fractional \emph{rms} at \emph{soft} X-ray energies is strongly linked to the progressive increase of a disc component, in analogy with what would be expected from a comparison with disc-dominated states.\\ 
In order to produce a drop of the fractional \emph{rms}, the disc component should be less variable than the power law.
Thus we estimated the fractional \emph{rms} intrinsic to the disc and the power law components.
To this aim we computed the \emph{broad-frequency-band} covariance spectra (with respect to the 0.5-10 keV reference band) of the XMM-Newton data (see Wilkinson \& Uttley 2009; Uttley et al. 2011; Uttley et al. 2014), which allow us to determine the fraction of variable disc emission. Note that, strictly speaking, the covariance spectrum gives the spectral shape of the components whose variability is linearly correlated with the variability in the reference band. However, for intrinsic coherence close to unity (as in this case, see also De Marco et al. 2015), the covariance spectrum is equivalent, though with smaller error bars (Wilkinson et al. 2009; Uttley et al. 2014), to the \emph{rms} spectrum, which gives the spectral shape of all the variable components. 
We fit both the covariance and the energy spectra with the same model used above, and with the $N_{H}$ parameter tied to the value obtained from the fit of the energy spectra. Fig. \ref{fig:covar} shows the ratios to the best-fit power law, once the normalization of the disc-component is set to zero. 
All the covariance spectra show an excess at soft X-ray energies. This demonstrates the presence of a variable disc component at all the luminosities spanned by the analysed hard-state observations. The intensity of the excess in the covariance spectra is lower than in the energy spectra in the medium and high-luminosity observations, indicating that some fraction of the total disc emission is constant or that the disc is less variable. 
We used the best-fit models to derive estimates of the intrinsic fractional \emph{rms} of the disc and the power law components (for each of the two components this is computed as the ratio between the absorbed flux in the covariance and in the energy spectrum). 
These are reported in Table \ref{tab:rms_comp}. In addition, we report the fractions of variable/constant power law/disc flux contributing to the total \emph{soft} band flux.

We observe that the disc is intrinsically variable at all luminosities. Its fractional \emph{rms} is comparable with that intrinsic to the power law component (at low luminosities it is consistent with being as variable as the power law), and much higher than typically observed in soft states (a few percents; Mu\~noz-Darias et al. 2011).
However, the disc fractional \emph{rms} decreases with luminosity faster than the power law fractional \emph{rms}. Combined with the variation of the relative contribution from the disc and the power law to the total \emph{soft} band flux (Fig. \ref{fig:rms_vs_disc}), the result is a net increase (by a factor of $\sim$4) of constant disc emission in the \emph{soft} band ($D_{const}$) as a function of luminosity. In comparison, the constant power law emission in the \emph{soft} band ($P_{const}$) increases only by a factor $\sim$1.6. 
Thus we conclude that the drop of fractional \emph{rms} in the \emph{soft} band is driven by the net increase of constant disc flux in this band. It is worth stressing that all these considerations refer to the \emph{broad-frequency-band}, while a more detailed treatment of the \emph{high-frequency} (i.e. above the peak-frequency of the high-frequency Lorentzian) PSD properties is deferred to Sect. \ref{sec:scaling}.

\begin{table}
\caption{The fractional \emph{rms} of the disc (in the \emph{soft} band) and the power law (in both the \emph{soft} and the \emph{hard} band) components, and the fractions of the total \emph{soft} band flux due to variable/constant disc/power law emission ($D_{var}$, $P_{var}$ and $D_{const}$, $P_{const}$). These estimates have been obtained by fitting the \emph{broad-frequency-band} (0.002-20 Hz, 0.017-20 Hz, and 0.04-20 Hz, respectively for the low, medium, and high luminosity observations) covariance spectra of GX 339-4 and its XMM-Newton energy spectra with the model described in Sect. \ref{sec:spec}. All these estimates have been derived considering the absorbed fluxes. Errors are reported at 90 percent confidence level.}
\label{tab:rms_comp}
\centering
\vspace{0.2cm}

\begin{tabular}{c c c c }
\hline          

              &  Low L   &  Med L      &  High L \\
\hline

& \multicolumn{3}{ |c| }{Disc [0.5-1.5 keV]} \\
\hline
 fractional \emph{rms}  &   0.65$\pm$0.20  & 0.40$\pm$0.03   &   0.20$\pm$0.03  \\                
 $D_{var}$      & 0.17$\pm$0.03  &    0.15$\pm$0.01   & 0.09$\pm$0.01   \\
 $D_{const}$   & 0.09$\pm$0.08  & 0.24$\pm$0.01 &  0.36$\pm$0.02 \\
 \hline
 & \multicolumn{3}{ |c| }{Power law [0.5-1.5 keV]} \\
 \hline
 fractional \emph{rms}  & 0.68$\pm$0.02     &    0.47$\pm$0.01    &  0.34$\pm$0.01  \\ 
 $P_{var}$      &  0.51$\pm$0.04  & 0.29$\pm$0.01 &  0.19$\pm$0.01 \\
 $P_{const}$   &  0.23$\pm$0.03 & 0.32$\pm$0.01 & 0.37$\pm$0.01  \\
 \hline
   & \multicolumn{3}{ |c| }{Power law [2-9 keV]} \\
   \hline
 fractional \emph{rms} & 0.59$\pm$0.02       &   0.42$\pm$0.01    &  0.33$\pm$0.01  \\ 
 \hline

\end{tabular}

\end{table}

\begin{figure}
\centering
\vspace{1.5cm}
\begin{tabular}{p{8cm}}

\includegraphics[height=6cm,angle=0]{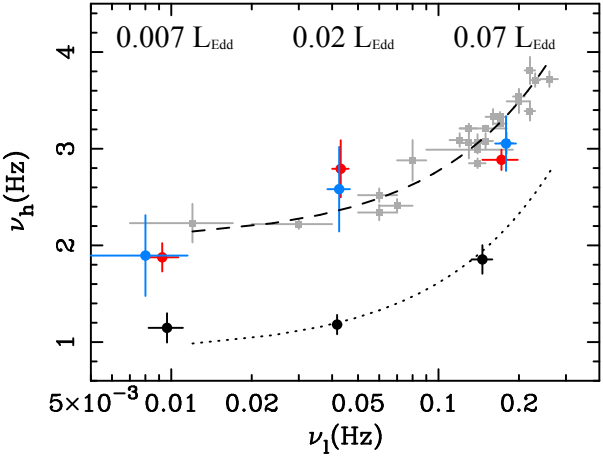} 

 \end{tabular}
\caption{The best-fit peak frequencies of the high- and low-frequency Lorentzians. The colors refer to the energy bands: \emph{soft} (black), \emph{hard} (red), and \emph{very hard} (blue). The grey points are the frequencies of the high- and low-frequency Lorentzian components as obtained by Belloni et al. (2005) from the analysis of all the hard-state RXTE observations of GX 339-4 during the 2002/2003 outburst. The dashed curve is a linear best-fit to the Belloni et al. (2005) data points, while the dotted curve is the same linear model (with only the intercept left free to vary) fit to the \emph{soft} band data points derived in this paper.}
\label{fig:cont}

\end{figure}

\subsection{Testing the correlation between the characteristic frequencies of the PSD in the \emph{soft} band}
\label{sec:pfreq}

\begin{figure}
\centering
\vspace{1.5cm}
\begin{tabular}{p{8.8cm}}

\includegraphics[width=0.85\textwidth,angle=270]{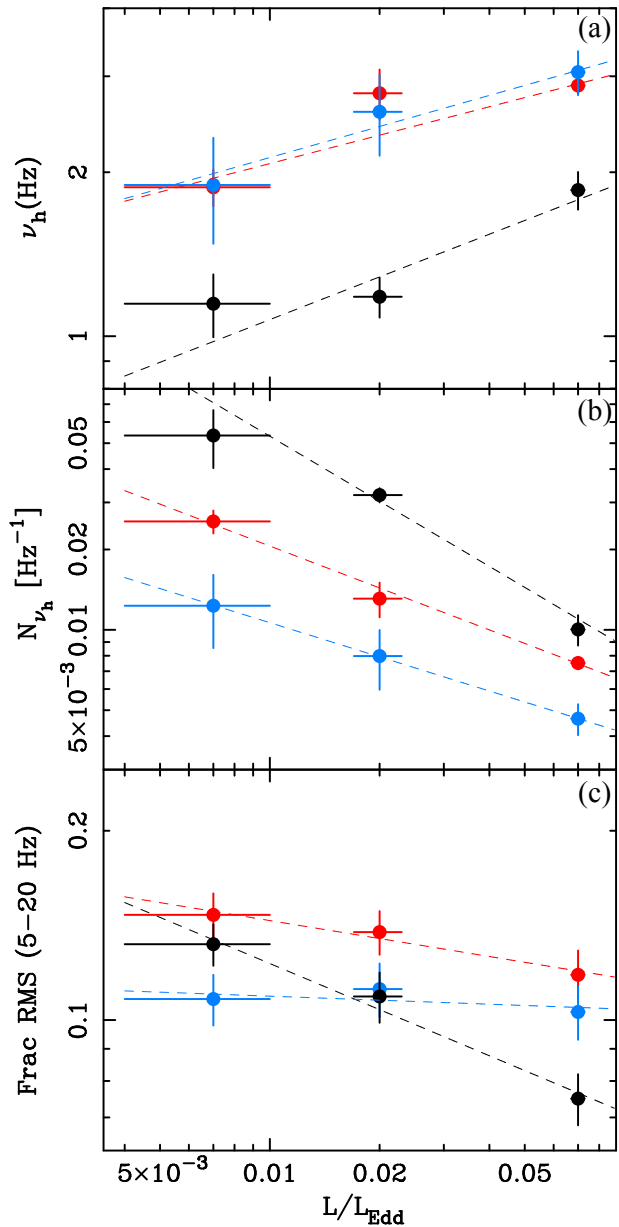} 

 \end{tabular}
\caption{The scaling as a function of $L/L_{Edd}$ (where $L$ is the 3-30 keV luminosity) of: \emph{(a)} the peak-frequency of the high-frequency Lorentzian, $\nu_{h}$; \emph{(b)} the normalization of the PSD at the peak-frequency of the high-frequency Lorentzian, $N_{\nu_{h}}$; \emph{(c)} the \emph{high-frequency} (5-20 Hz) fractional \emph{rms}. The dashed lines are the best-fit models. In the plots, the colors refer to the energy bands: \emph{soft} (black), \emph{hard} (red), \emph{very hard} (blue).}
\label{fig:nuvsL}

\end{figure}

Fig. \ref{fig:cont} reports the best-fit peak frequency of the low-frequency Lorentzian, $\nu_{l}$, plotted against the high-frequency one, $\nu_{h}$ (these two frequencies are marked by arrows in the plots of Fig. \ref{fig:PSD1}). 
The peak frequency of each Lorentzian is computed using the formula (e.g. Belloni et al. 2002):
$$
\nu=\sqrt{\nu_0^2 + \Delta^2}
$$
where $\nu_0$ is the centroid frequency and $\Delta$ is the half-width-at-half-maximum (HWHM) of the Lorentzian. Note that flat-top, broad-band noise components are best-fit by Lorentzians with $\nu_0$ consistent with zero. This is the case for most of our fits. However, in some cases, the high-frequency Lorentzian fit required a more peaked profile, with $\nu_0$ different from zero.
The error on the peak frequency is computed from the 90 percent confidence contours of the $\nu_0$ and $\Delta$ parameters. Both $\nu_{l}$ and $\nu_{h}$ show the characteristic shift towards high frequencies as a function of luminosity. While $\nu_{l}$ shifts by a factor of $\sim$20, $\nu_{h}$ shifts by a factor of $\sim$1.6, for a variation in luminosity of a factor $\sim10$. The black dots show that this same shift is observed also in the \emph{soft} energy band. In addition we observe that $\nu_{h}$ is systematically and significantly (at $>3\sigma$ confidence level) offset towards lower frequencies in the \emph{soft} band. 
Note that this offset is driven by a variation of $\Delta$, indicating that the high-frequency Lorentzian is systematically narrower in the \emph{soft} band, than at higher energies. It is also worth noting that, if the high-frequency PSDs (e.g. 1.5 times above $\nu_h$) are fitted with a simple power law, a significant steepening of the slope is observed in the \emph{soft} band of the high-luminosity observation. This resembles the behaviour observed in AGN in the disc-dominated optical band (e.g. Mushotzky et al. 2011).
On the other hand, $\nu_{l}$ is consistent with remaining constant among the different energy bands of the same observation. Indeed, for each luminosity, the low-frequency Lorentzians in the three energy bands are co-aligned in the plots of Fig. \ref{fig:PSD1}, while an offset of $\nu_{h}$ in the \emph{soft} band is clearly observed (see also Fig. \ref{fig:nuvsL}, panel $a$, and Sect. \ref{sec:scaling} for a description of this figure). Note that the peak frequency of the additional third Lorentzian does not show any significant energy-dependence (see also Table \ref{tab:fits}). These results suggest that the \emph{high-frequency} PSD behaves in a different manner in the \emph{soft} band than at higher energies. This aspect will be explored in more detail in Sect. \ref{sec:scaling}.\\
In order to compare our results with those reported in the literature, we overplot in Fig. \ref{fig:cont} (grey points) the characteristic frequencies associated with the high- and low-frequency Lorentzians (in the energy band $E=3.8-15.3$ keV) as obtained by Belloni et al. (2005) from the analysis of all the RXTE observations of the 2002/2003 outburst of GX 339-4 (only the values relative to the hard state of the source are shown in the plot). 
The two main Lorentzians of the \emph{hard}- and \emph{very hard}-band PSDs analysed in this paper (red and blue points in Fig. \ref{fig:cont}) follow the same trend and span the same range of characteristic frequencies as those of the 2002/2003 RXTE data. This trend seems to characterize also the Lorentzians in the \emph{soft}-band PSD, despite the significant offset.
To show that the correlation between $\nu_{l}$ and $\nu_{h}$ is preserved in the \emph{soft} band, we fit the Belloni et al. (2005) data points with a linear model (in the log-linear plot of Fig. \ref{fig:cont} the model is represented by the dashed curve). Then we fixed the slope to the best-fit value, and letting the intercept free to vary, we fit the \emph{soft} band data points (dotted curve in Fig. \ref{fig:cont}). Though the number of data points is small, we obtain a very good description of the $\nu_{l}$ vs. $\nu_{h}$ trend in the \emph{soft} band, thus in agreement with the idea that the two Lorentzians are linked by the same linear relation as observed at high energies. According to these fits, the offset of $\nu_{h}$ as measured in the \emph{soft} band and at higher energies is of a factor $\sim 2$.

\subsection{The \emph{high-frequency} PSD of GX 339-4}
\label{sec:scaling}

We now focus on the \emph{high-frequency} PSD of GX 339-4 (between $5-20$ Hz, i.e. above the peak of the high-frequency Lorentzian).
In the analysed datasets we observe that the peak of the high-frequency Lorentzian varies with luminosity by a factor of $\sim$1.6 (Sect. \ref{sec:pfreq}), as also shown in Fig.~\ref{fig:nuvsL} (panel $a$), where we plot $\nu_{h}$ as a function of $L/L_{Edd}$ (note that $L$ is computed in the 3-30 keV range). A fit with a linear model in the log-log space returns a best-fit slope of $0.26\pm0.11$, $0.17\pm0.06$, and $0.19\pm0.16$, respectively in the \emph{soft}, \emph{hard}, and \emph{very hard} band. We note that these slopes are consistent with being the same, but consistently and significantly different from a slope of 1. However the \emph{soft} band has a significantly lower intercept (differing by a factor $\sim0.7\pm0.2$), this being a 
consequence of the energy-dependence of $\nu_{h}$ (Sect. \ref{sec:pfreq}).

We also looked at the variations of the normalization of the PSD at the peak-frequency of the high-frequency Lorentzian ($N_{\nu_{h}}$, in units of $Hz^{-1}$) as a function of luminosity. 
We derived $N_{\nu_{h}}$ and its related uncertainties (at the 90 percent confidence level) from the best-fit PSD models. Results are shown in Fig. \ref{fig:nuvsL} (panel $b$). 
A fit with a linear model in the log-log space gives best-fit slopes of $-0.81\pm0.13$, $-0.52\pm0.08$, and $-0.42\pm0.20$, respectively in the \emph{soft}, \emph{hard}, and \emph{very hard} band (dashed lines in Fig. \ref{fig:nuvsL}, panel $b$). 
These results point to the existence of an inverse-power scaling relation between $N_{\nu_{h}}$ and $L/L_{Edd}$. 
Once again we note a consistent behaviour between the \emph{hard}, and \emph{very hard} bands, and a significant deviation in the \emph{soft} band.
Indeed, in the \emph{soft} band this trend appears steeper than in the other bands.

Finally, we computed the fractional \emph{rms} in the $5-20$ Hz frequency range. Results are reported in Table \ref{tab:rms} and plotted in Fig. \ref{fig:nuvsL} (panel $c$).
Despite the significant variations of the high-frequency Lorentzian as a function of luminosity, the \emph{high-frequency} fractional \emph{rms} (above the peak-frequency of the high-frequency Lorentzian) in the \emph{hard} and \emph{very hard} bands is consistent with being remarkably constant. This is due to the combined effect of the frequency shift \emph{and} the decrease of normalization of the high-frequency PSD with luminosity. These combined variations produce a constant \emph{high-frequency} fractional \emph{rms}. Indeed, fitting the data with a constant we measure a scatter around this best-fit value of $\sim 10$ and $\sim 4$ percent (respectively in the \emph{hard} and \emph{very hard} band), i.e. of the order of the corresponding average uncertainty (Fig. \ref{fig:PSD2}, middle and lower panel).
On the other hand, the constancy of the \emph{high-frequency} fractional \emph{rms} with luminosity breaks down
in the \emph{soft} energy band (Fig. \ref{fig:PSD2}, upper panel). If fitted with a constant model, the 
 increases to $\sim 30$ percent. Instead, a fit with a linear model in the log-log space yields a best-fit slope of $-0.24\pm0.08$, and the scatter around the model is of $\sim$ 3 percent. This is due to the fact that the normalization of the high-frequency PSD decreases with luminosity faster in the \emph{soft} band than at higher energies.

\section{Discussion}
\label{discussion}

In this paper we investigated the luminosity-dependence of the characteristic frequencies and variability power of GX 339-4 in the hard state. These properties have been extensively studied in the past, through the use of RXTE data, which cover energies $>3$ keV. Our work aims at extending this analysis down to the \emph{soft} X-ray energy band ($E\geq 0.5$ keV), where the contribution of the disc thermal emission becomes non-negligible. For this reason we used XMM-Newton and simultaneous RXTE observations, covering the energy band 0.5-30 keV. 
The selected observations belong to two different outbursts of the source. Nonetheless, the variability properties of BHXRBs are extremely similar from outburst to outburst (Mu\~noz-Darias et al. 2011, Grinberg et al. 2014), thus justifying our approach of considering observations taken during different activity periods.

\subsection{Disc variability and the shape of the \emph{broad-frequency-band} fractional \emph{rms} spectra}
\label{sect:shift}

The drop of fractional \emph{rms} that characterizes the transition from the hard to the soft state, is ascribed to the emergence of a disc component, which, in the soft state, is more intense and less variable (with typical fractional \emph{rms} values of few percents) than the power law. The fractional \emph{rms} spectrum is observed to change accordingly, from an ``inverted'' (in the hard state) to a ``hard'' (in the soft state) spectral shape.

Extending our analysis of the hard-state variability properties of GX 339-4 down to the \emph{soft} X-ray band, we observe a drop of fractional \emph{rms} that correlates with the progressive increase of disc flux relative to the total flux in that band. This resembles the behaviour seen in soft states, although to a much lower extent (indeed the disc fractional \emph{rms} is much higher and the drop less intense than in soft states, see inset of Fig. \ref{fig:rms}), and is expected if the disc has lower fractional \emph{rms} than the power law.
This raises the question as to whether or not the disc in the hard state has similar timing properties as in the soft state. 
Analysing the energy and covariance spectra of GX 339-4, we observe intrinsic \emph{broad-frequency-band} disc variability at all the luminosities spanned by the hard-state observations.
The presence of a variable disc in a hard state observation (corresponding to our medium luminosity observation) was first reported by Wilkinson \& Uttley (2009). 
We detect high levels of disc variability ($\sim20-65$ percent) in the three observations, comparable to the variability intrinsic to the power law component, and significantly higher than observed in soft states.
However, the disc fractional \emph{rms} decreases with luminosity faster than the power law fractional \emph{rms}. At the same time, the fraction of \emph{soft} band flux due to the disc (power law) component increases (decreases) with luminosity (Fig. \ref{fig:rms_vs_disc}). The net effect is a strong increase (a factor of $\sim$4) of constant disc-flux fraction in the \emph{soft} band, which causes the observed drop of fractional \emph{rms}. It is worth noting that, without the contribution from the disc component, the fractional \emph{rms} spectra always show an ``inverted'' shape (as typically seen in the RXTE data, e.g. Belloni et al. 2011), with variability power decreasing as a function of energy.
 We conclude that a relatively high fraction of disc flux is variable during the hard state of GX 339-4. 
However, the increasing fraction of constant disc flux with luminosity in the \emph{soft} band is responsible for the net drop of fractional \emph{rms} as compared to power law-dominated energy bands. We argue that this trend continues throughout the outburst, thus leading to the disc being mostly constant during the soft state, and significantly different from the disc seen in hard states.

\subsection{The scaling of the high-frequency PSD with luminosity in BHXRBs and AGN}
\label{discuss:highfvar}

A number of correlations link the \emph{high-frequency} variability properties of BHXRBs and AGN, suggesting that the dominant variability process at these frequencies is the same in both the classes of objects.
In particular, considering a sample of 10 AGN and 2 BHXRBs in the soft state, McHardy et al. (2006) found a linear scaling of the high-frequency break (a characteristic frequency of AGN PSDs, Edelson \& Nandra 1999, commonly associated with $\nu_{h}$ in BHXRBs) with the BH mass ($M_{BH}$) and the mass-accretion rate ($\dot{m}_{Edd}$, i.e. normalized by the mass-accretion rate required to reach the Eddington luminosity), which defines the so-called `variability plane'. Koerding et al. (2007) found a similar scaling relation also for sources in the hard state (but offset with respect to the corresponding relation valid for soft-state sources).

As a consequence, we would expect to observe $\nu_{h}$ scaling linearly with the mass-accretion rate also in the hard-state observations of GX 339-4 analysed in this paper. However, we see $\nu_{h}$ scaling with $L/L_{Edd}$ (with $L=L_{3-30 keV}$), with an average slope of $\sim 0.21\pm 0.11$, therefore not consistent with the results by McHardy et al. (2006) and Koerding et al. (2007)\footnote{Here we assume $L_{bol}\propto L$ and $L_{bol}/L_{Edd}\propto\dot{m}_{Edd}^{\beta}$, with $\beta=1-2$, the exact value depending on the accretion efficiency of the flow (Fender \& Mu\~noz-Darias 2015).}.
We note, however, that our result is in agreement with the recent estimate obtained from the PSD analysis of a sample of 104 nearby AGN observed with XMM-Newton (Gonzalez-Martin \& Vaughan, 2012), which reports a best-fit slope of $0.24\pm0.28$. 
This therefore suggests that, indeed, the same engine is at work in both AGN and BHXRBs, but that the dependence of the \emph{high-frequency} variability on luminosity/mass-accretion is significantly shallower than previously thought (McHardy et al. 2006; Koerding et al. 2007).
We also investigated the dependence of the normalization of the PSD at the break, $N_{\nu_{h}}$, on the luminosity. 
We find that $N_{\nu_{h}}$ decreases as a function of $L/L_{Edd}$. 
Interestingly, this trend can counterbalance the $\nu_{h}$ vs. $L/L_{Edd}$ scaling, producing the same \emph{high-frequency} variability power at different luminosities.

Previous studies have pointed out that the \emph{high-frequency} PSD does not show a clear scaling with the luminosity/mass-accretion rate both in BHXRBs (Gierli\'nski et al. 2008) and AGN (O'Neil et al. 2005; Zhou et al. 2010; Ponti et al. 2012).
We observe that the \emph{high-frequency} fractional \emph{rms} of GX 339-4 is constant in the \emph{hard} and \emph{very hard} band (dominated by the hard X-ray Comptonization continuum) at any luminosity probed. In particular we find that this quantity has a small scatter ($4-10$ percent, in agreement with results by Mu\~noz-Darias et al. 2011 and by Heil, Vaughan, \& Uttley 2012) in these two energy bands. The remarkable constancy of the \emph{high-frequency} PSD provides us with a tool to measure the BH mass of accreting objects (Niko{\l}ajuk et al. 2004; Gierlinski et al. 2008; Ponti et al. 2012; Kelly et al. 2011; La Franca et al. 2014). Indeed the \emph{high-frequency} fractional \emph{rms} (or equivalently the `excess variance', e.g. Niko{\l}ajuk et al. 2004; Ponti et al. 2012), while being independent of the mass-accretion rate, scales inversely with the BH mass, with a very small scatter (which mostly depends on the uncertainty on $M_{BH}$), of the order of less than 0.2-0.4 dex (e.g. O'Neill et al. 2005; Zhou et al. 2010; Ponti et al. 2012; Kelly et al. 2011).

As observed in Sect. \ref{sec:scaling}, the dependences of $\nu_{h}$ and $N_{\nu_{h}}$ are significantly different in the \emph{soft} band. In particular, the constancy of the \emph{high-frequency} fractional \emph{rms} as a function of luminosity breaks down at low energies, with a scatter around the best-fit constant model increasing to $\sim$ 30 percent.
We ascribe this to the presence of the additional contribution from the disc component. Interestingly the same behaviour is observed in AGN. Indeed, it is observed that the scatter in the relation between the short-time scale variability and the BH mass is larger in the optical band (dominated by the disc emission) than in the X-ray band (dominated by the power law), and the optical variability of AGN is significantly anticorrelated with the optical luminosity (Kelly et al. 2013).

\subsection{A possible interpretation for the offset of the high-frequency Lorentzian in the \emph{soft} band}
\label{sec:offset}
The characteristic frequencies in the PSD of accreting systems (i.e. the peak-frequency of each Lorentzian) are likely related to fundamental time scales of the accretion flow at specific radii (e.g. Psaltis, Belloni \& van der Klis 1999; Churazov, Gilfanov, \& Revnivtsev 2001; Motta et al. 2014). 
Though not yet conclusive, the general picture that emerged over the years is that the low-frequency Lorentzian in the hard state of BHXRBs is associated with a characteristic time scale (most probably the viscous time scale) at the inner edge of the accretion disc (e.g. Churazov et al. 2001; Done et al. 2007).
According to recent studies, variations intrinsic to the seed-photons flux produce the long-time scale variability (i.e. the low-frequency Lorentzian) observed in the \emph{soft} energy band (Sect. \ref{sec:spec}; Wilkinson \& Uttley 2009; Uttley et al. 2014). These variations are due to perturbations in the accretion rate, which travel inward and modulate the hard X-ray emission from the inner corona, thus producing the long-time scale variability observed also in harder energy bands (e.g. Lyubarskii 1997; Kotov, Churazov, \& Gilfanov 2001; Ar\'evalo \& Uttley 2006; Ingram \& van der Klis 2013).
The existence of additional variability components at frequencies higher than those corresponding to the inner edge of the disc, can be explained provided the corona extends down to the innermost stable circular orbit (ISCO, Motta et al. 2014) and is intrinsically variable on short time scales (e.g. Churazov, Gilfanov, \& Revnivtsev 2001). This inner hot flow/corona is thus responsible for producing the high-frequency Lorentzian observed in \emph{hard} and \emph{very hard} energy bands.\\
Such a high-frequency component is observed also in the \emph{soft} band, and is linearly correlated with that in harder energy bands (as seen from covariance spectra, light grey triangles in Fig. \ref{fig:covar}; see also Uttley et al. 2011; De Marco et al. 2015). We showed that the high-frequency Lorentzian is significantly offset, by a factor of $\sim$2, towards lower frequencies in the \emph{soft} band with respect to higher-energy bands. 
The covariance spectra (light grey triangles in Fig. \ref{fig:covar}) show that the high-frequency PSD contains significant contribution from variable disc emission, so the offset, \emph{soft} band high-frequency Lorentzian is likely linked to variability coming from the disc component.
However, within the scenario depicted above, intrinsic disc variability due to accretion rate perturbations occurring on time scales shorter than the viscous time scale at the inner edge of the disc is expected to be highly suppressed.
Thus, a possible explanation is that this high-frequency variability is not intrinsic to the disc, but results 
from reprocessing of the short-time scale variable corona emission (e.g. thermal reprocessing; Wilkinson \& Uttley 2009; Uttley et al. 2011; De Marco et al. 2015).
In standard disc-corona geometries, the hard X-rays would irradiate a relatively \emph{extended} region of the disc. 
Thus, the reprocessing region would act as a low-pass filter, damping the high-frequency variability power in the \emph{soft} band (e.g. Ar\'evalo \& Uttley 2006; Uttley et al. 2014). As a consequence the \emph{soft} band high-frequency Lorentzian would be narrower, and its peak-frequency would result slightly shifted towards low frequencies with respect to the corresponding peak in harder energy bands. This resembles the behaviour observed in the data of GX 339-4.  
According to this interpretation, the relative offset of the peak-frequency of the high-frequency Lorentzian in the different bands should trace the variations of the geometry of the accretion disc/corona during the outburst.

\section*{Acknowledgments}

This work is based on observations obtained with XMM-{\it Newton}, an ESA science mission with instruments and contributions directly funded by ESA Member States and NASA. GP acknowledges support via an EU Marie Curie Intra-European fellowship under contract no. FP-PEOPLE-2012- IEF-331095. This project was funded in part by European Research Council Advanced Grant 267697 4-$\pi$-sky: Extreme Astrophysics with Revolutionary Radio Telescopes. The authors thank the anonymous referee for helpful comments which allowed to significantly improve the paper.

\end{document}